\documentclass[12pt]{article}
\usepackage[dvips]{graphicx}
\usepackage{pdproc}

  \makeatletter 
  \def\@cite#1{[#1]} 
  \makeatother    
\begin{document}

\renewcommand{\thefootnote}{\alph{footnote}}

\title{
Extracting Higgs Boson Couplings from LHC Data
}

\author{ HEATHER E.\ LOGAN}

\address{ 
Department of Physics, University of Wisconsin\\
1150 University Avenue, Madison, Wisconsin 53706 USA
\\ {\rm E-mail: logan@physics.wisc.edu}}

\abstract{
We show how LHC Higgs boson production and decay data can be used to
extract the gauge and fermion couplings of the Higgs boson.
Incomplete input data leads to parameter degeneracies, which can be
lifted by imposing theoretical assumptions.  We show how successive
theoretical assumptions affect the parameter extraction, starting with
a general multi-Higgs doublet model and finishing with specific
supersymmetric scenarios.
}

\normalsize\baselineskip=15pt

\section{Higgs Couplings at the LHC}

The LHC will be able to observe a 
Higgs boson in a variety of production and decay channels, especially 
in the intermediate mass range 114 GeV $< m_H < 200$ GeV.
The event rates in these production and decay channels are 
determined by the Higgs couplings; thus, measurements of the rates in
multiple channels allow various combinations of Higgs couplings to be
determined.  These measurements can be combined to form an error ellipse
in the parameter space of Higgs couplings to each Standard Model (SM) 
species.

Unfortunately, not all possible Higgs decays can be directly observed
at the LHC; for example, decays to gluon pairs or light quark jets are
buried under enormous backgrounds.  Likewise, no way is known to measure
any of the Higgs production cross sections independent of decay mode.
Because of this, 
there are large correlations among the individual couplings: the error
ellipse is elongated along various directions in the parameter space.
Projecting the error ellipse onto each coupling axis then gives quite
poor measurements of the individual couplings.

In order to lift the correlations, additional theoretical assumptions
are needed.  An earlier study \cite{ZKNR}
assumed no unexpected Higgs decay channels and fixed the ratio of Higgs
couplings to $b$ and $\tau$ to its SM value, allowing $H\to \tau\tau$ data to
fix the poorly-measured $H\to b \bar b$ coupling.  These assumptions 
allowed the Higgs total width to be extracted from the observed decay
modes, then used to solve for the individual couplings.  These assumptions
were quite restrictive; in particular, the ratio of $b$ and $\tau$ couplings
can differ from its SM value in the MSSM.  A later study \cite{BR}
removed the $b/\tau$ assumption by including in the fit a 
channel with $H \to b \bar b$.

In this talk, based on \cite{us}, we take a different strategy.  We assume
only that the Higgs couplings to $WW$ and $ZZ$ are bounded from above by their
SM values.  This is a mild assumption: it is true in a general 
multi-Higgs-doublet model, with or without additional singlets.  In particular,
it is true in the MSSM.

The theoretical constraint works as follows.
First, the mere observation of Higgs production puts a lower bound on the 
production couplings, which in turn puts a lower bound on the Higgs total 
width.  This is model-independent and does not depend on any theory input.
Second, Higgs production in weak boson fusion (WBF) with
decays to $WW$ or $ZZ$ measures $\Gamma^2_V/\Gamma_{\rm tot}$.  
Combined with the theoretically imposed upper bound on $\Gamma_V$, this
gives an upper bound on the Higgs total width $\Gamma_{\rm tot}$.
The interplay between these two constraints on $\Gamma_{\rm tot}$ provides
constraints on the remaining Higgs couplings.

As a second approach, one can perform $\chi^2$ fits of the Higgs production
and decay rates to a particular model.  Imposing a particular model
selects out a subspace of the coupling parameter space,
effectively taking a slice through the error ellipse and resulting
in tighter constraints from the same LHC data.  We show fits within
a particular MSSM scenario as an example.

\section{Inputs and Fitting Procedure}

The Higgs production and decay channels used in our fits are shown in 
Table~\ref{tab1} (for details see \cite{us,Michael}).
\begin{table}
\caption{Channels included in the fit (marked with an X).  Production
modes are gluon fusion (GF, which dominates inclusive Higgs production),
weak boson fusion (WBF), $WH$ and $ZH$ associated production, and $t \bar t H$
associated production.}
\begin{center}
\begin{tabular}{|c|ccccc|}
\hline
Production & \multicolumn{5}{c|}{Decay} \\
 & $ZZ^{(*)}$ & $WW^{(*)}$ & $\gamma\gamma$ & $\tau\tau$ & $b \bar b$ \\
\hline
GF & X & X & X & & \\
WBF & X & X & X & X & \\
$WH$ &  & X & X &   & \\
$ZH$ &  &   & X &   & \\
$t \bar t H$ &  & X & X &  & X \\
\hline
\end{tabular}
\end{center}
\label{tab1}
\end{table}
These give us access to the Higgs couplings to gluon pairs, $W$ and $Z$ 
pairs, photon pairs, taus, $b$ quarks, and top quarks.
We take into account a large number of systematic uncertainties, including
the luminosity normalization, detection efficiencies, and theoretical 
uncertainties on production cross sections; see \cite{us} for details.
We perform fits within three LHC luminosity scenarios:
1) low luminosity, 30 fb$^{-1}$;
2) high luminosity, 300 fb$^{-1}$;
and 3) a mixed scenario with 300 fb$^{-1}$, of which only 100 fb$^{-1}$
are usable for WBF channels.  In all cases, we combine the statistics
from the two detectors, effectively doubling the delivered machine 
luminosity.

All channels listed in Table~\ref{tab1} have been studied at low luminosity,
and all channels except WBF have been studied at high luminosity.  The WBF
channels could suffer from high luminosity running because underlying
events could degrade the efficiency of their forward jet tag and minijet 
veto.  The mixed luminosity scenario is included to allow 
for such a degradation.
For the mixed and high luminosity scenarios, we scale up the signal and
background event numbers for WBF channels from the low luminosity studies.

\section{Results of the Fits}

We begin with the general fit valid in a multi-Higgs-doublet model, with or 
without additional singlets.
First, we assume $g^2_{W,Z} < 1.05 (g^2_{W,Z})_{\rm SM}$.  The extra 5\% margin
allows for theoretical uncertainties in the translation between 
couplings-squared and partial widths and also allows for small admixtures
of exotic Higgs states, such as SU(2) triplets.
Second, we allow for the possibility of additional particles running in
the loops for $H \to \gamma\gamma$ and $gg \to H$, fitted by a positive or
negative new partial width.
Finally, we allow for additional unobservable decays, such as to light 
hadrons, fitted with a partial width for unobservable decays.\footnote{Note
that truly invisible Higgs decays, such as to pairs of lightest neutralinos, 
could still be observed at the LHC through missing energy signals in 
WBF \cite{EZ}.}

\begin{figure}[htb]
\begin{center}
\resizebox{\textwidth}{!}{
\includegraphics{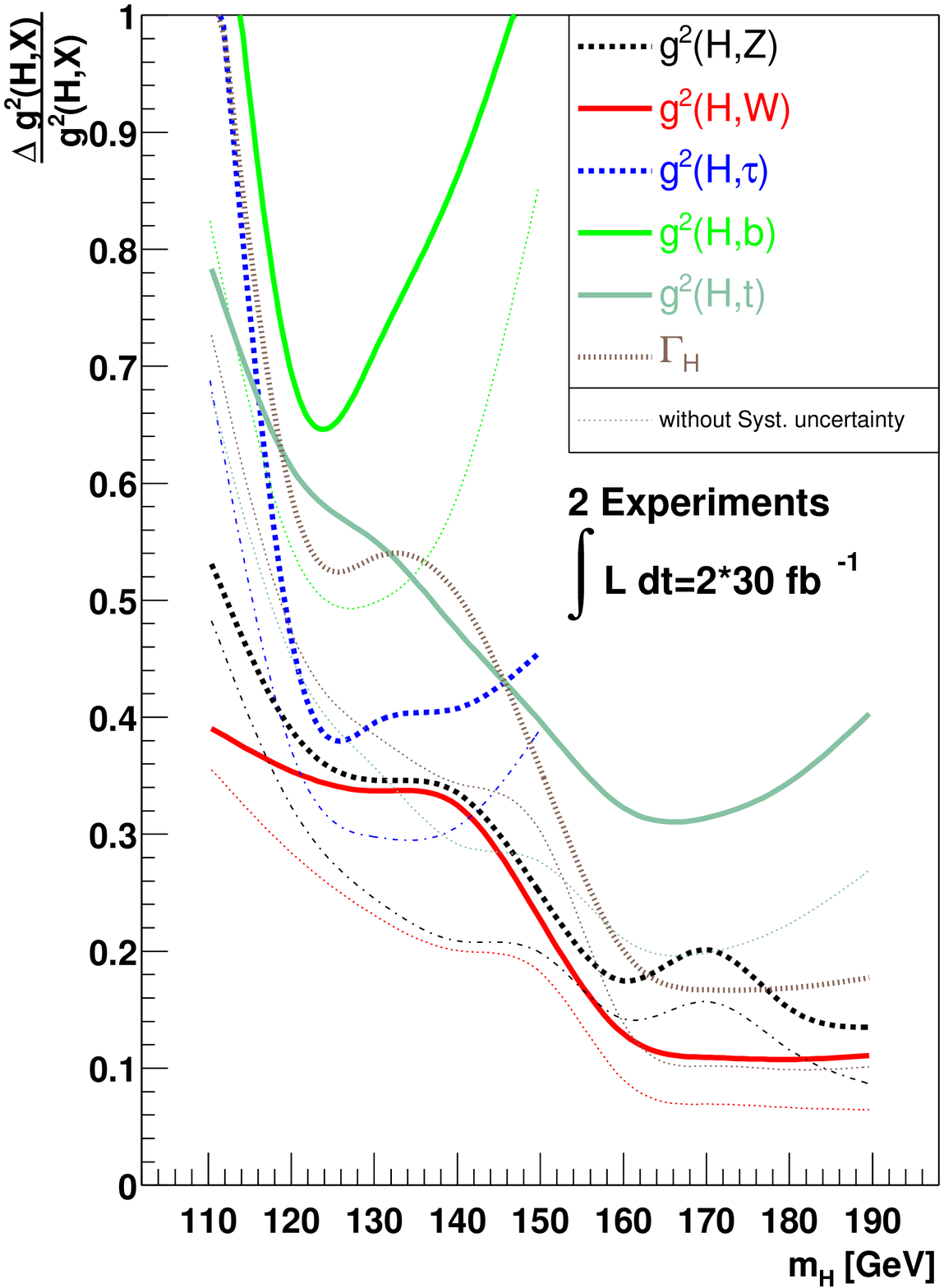}
\includegraphics{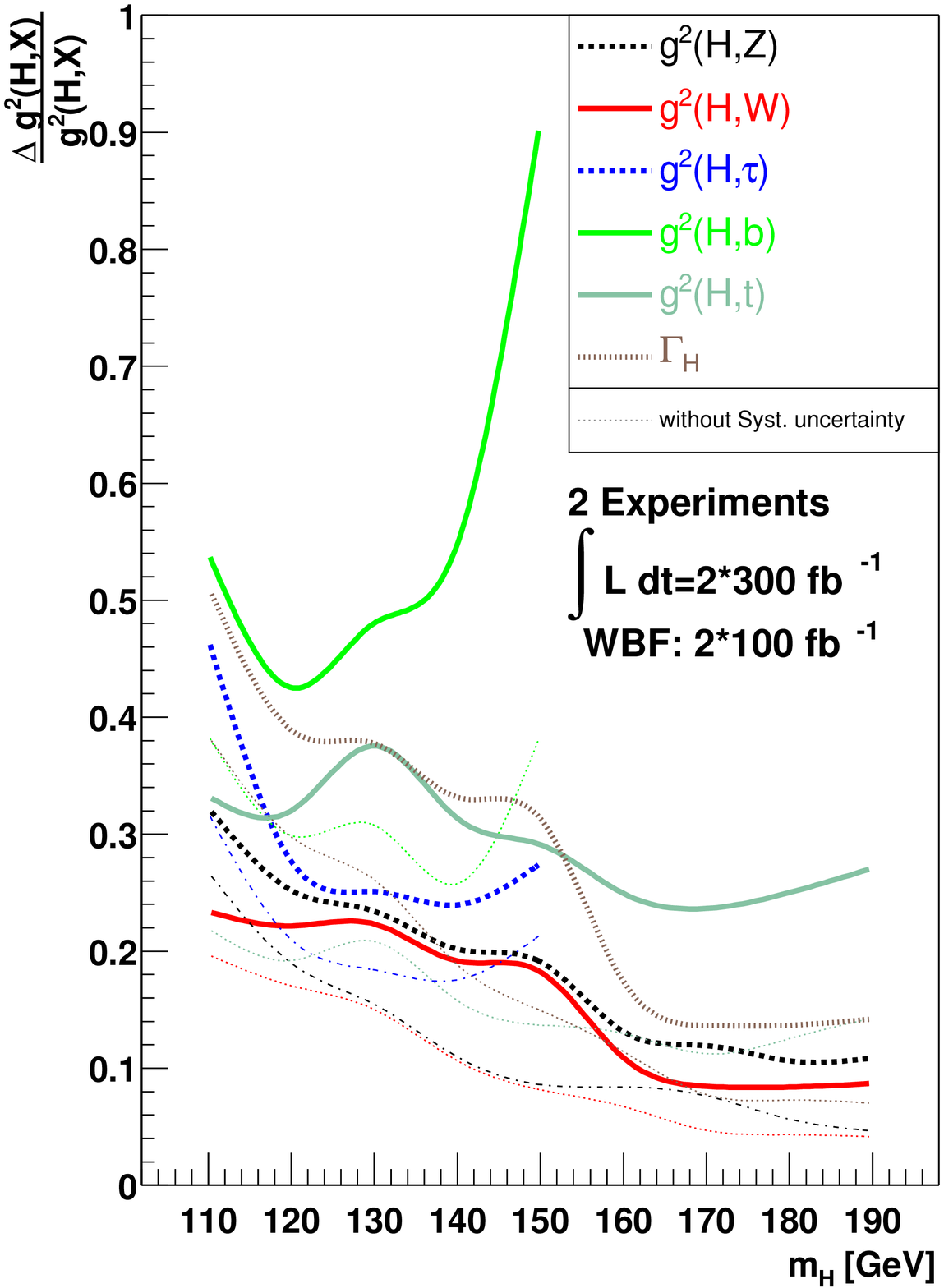}
}
\caption{%
Relative precisions of fitted Higgs couplings-squared as a function of
$m_H$ assuming SM rates in the low (left) and mixed (right) luminosity 
scenarios.  The precisions in the high luminosity scenario are about the
same as in the mixed scenario, except that the $\tau$ coupling extraction
improves.
}
\label{fig1}
\end{center}
\end{figure}

The results of the fit are shown in Fig.~\ref{fig1}.
In the mixed scenario the precision on the $W$ and $Z$ couplings-squared can 
reach 10\% at $m_H > 160$ GeV and about 25\% at lower $m_H$.
The precision on the top and $\tau$ couplings-squared is around 30\%
and that of the bottom coupling-squared reaches a minimum of 40\%.

\begin{figure}[htb]
\begin{center}
\resizebox{\textwidth}{!}{
\rotatebox{270}{\includegraphics[50,50][555,590]{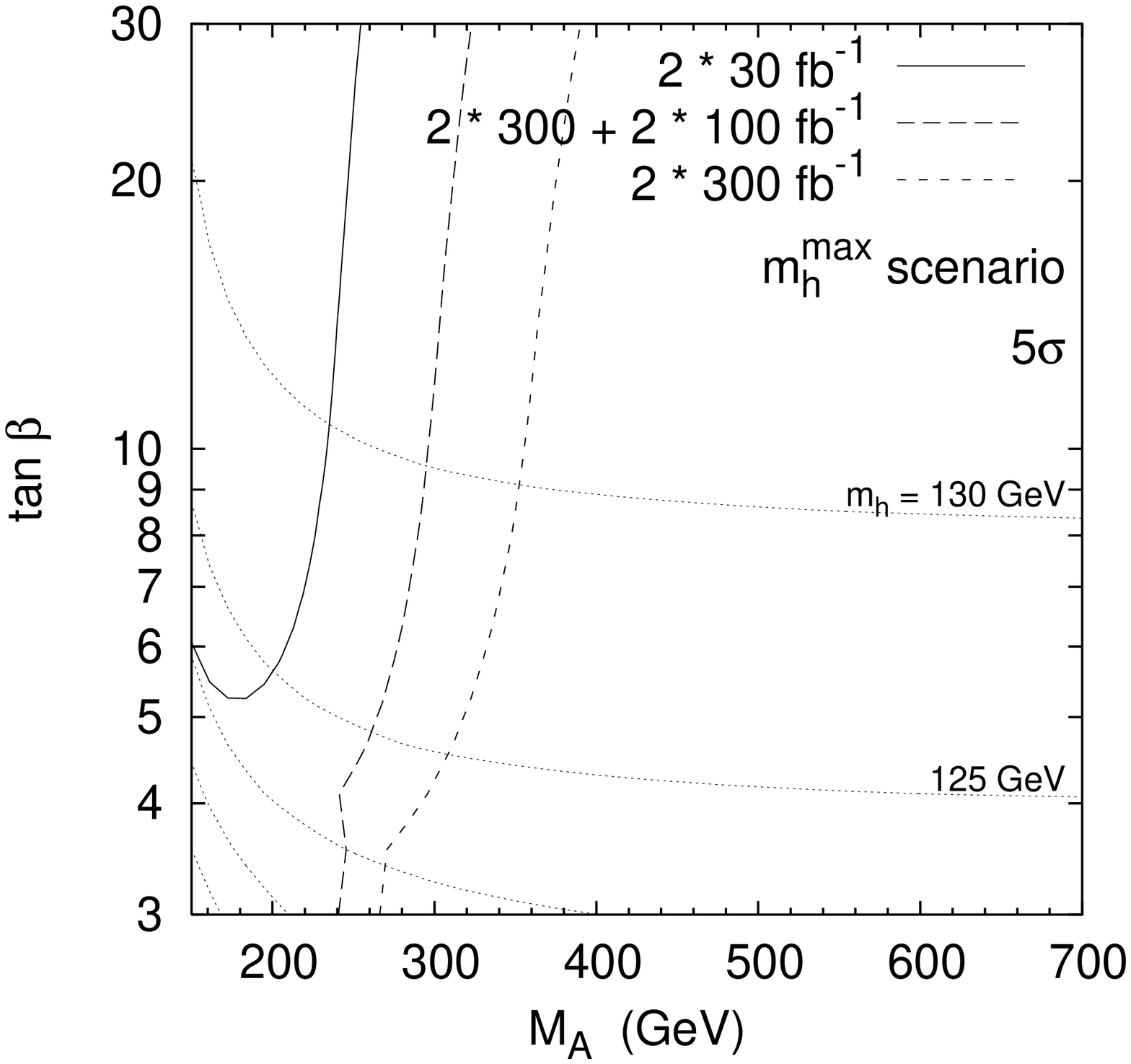}}
\rotatebox{270}{\includegraphics[50,50][555,590]{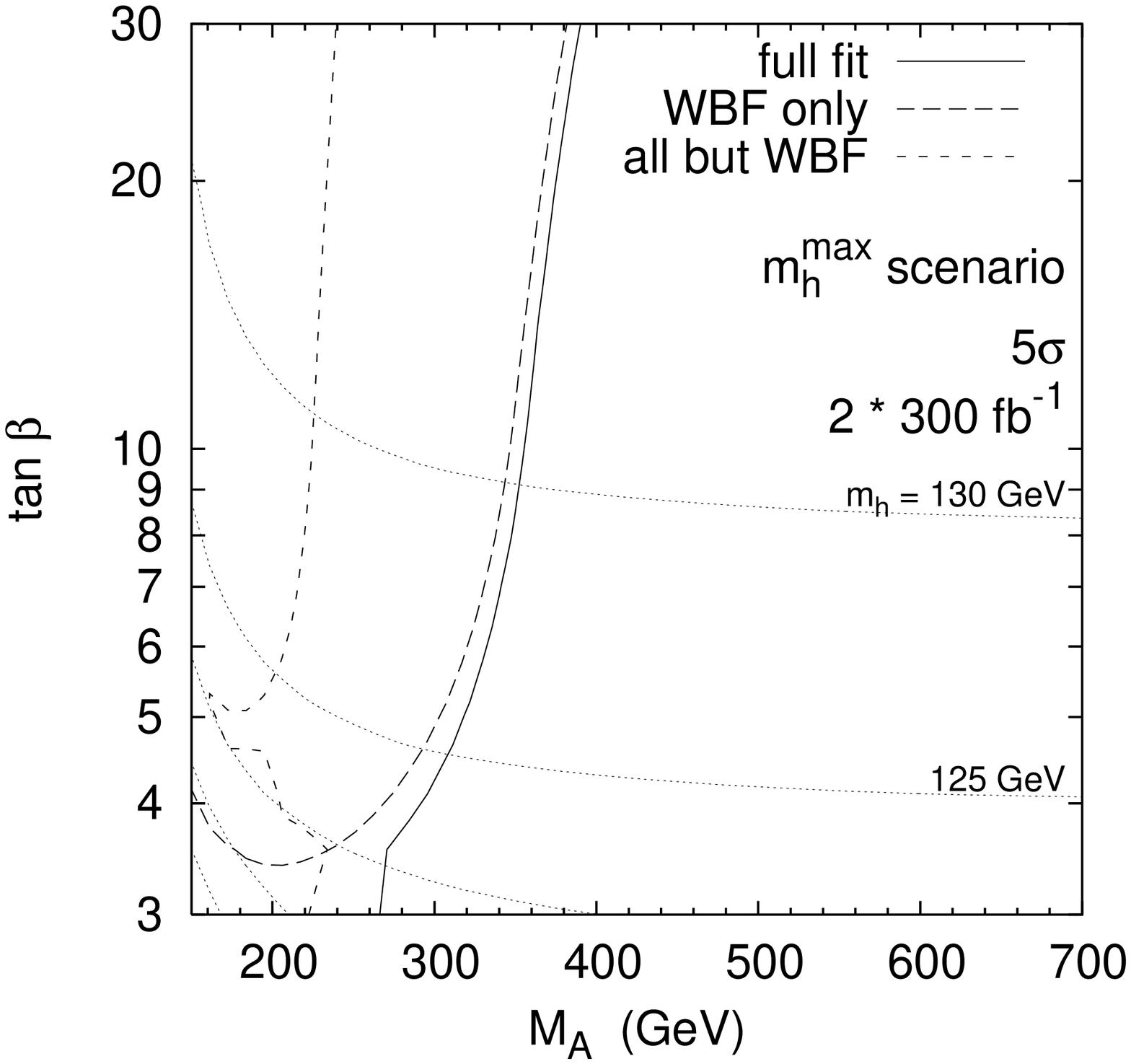}}
}
\caption{%
Fit within the MSSM $m_h^{\rm max}$ scenario in the $M_A$--$\tan\beta$
plane.  The region to the left of the curves would yield a 
$\geq 5 \sigma$ ($\Delta \chi^2 \geq 25$) discrepancy from the SM.
The mostly-horizontal dotted lines are contours of $m_h$ in steps of
5 GeV.
(Left) $\Delta \chi^2 = 25$ contours in the low (solid line), 
mixed (long-dashed), and high (short-dashed) luminosity scenarios.
(Right) $\Delta \chi^2 = 25$ contours in the high luminosity scenario 
using WBF channels only (long-dashed), all channels except WBF (short-dashed),
and the full fit (solid line).
}
\label{fig2}
\end{center}
\end{figure}

Within a particular model, we can ask whether a deviation from the SM
is visible in the Higgs boson couplings.
In the MSSM, the properties of the lightest Higgs boson $h$ approach those
of the SM Higgs in the decoupling limit of large $M_A$. We perform a 
$\chi^2$ fit in particular MSSM scenarios to see how far into the decoupling
regime $h$ can be distinguished from the SM.  Results are shown in 
Fig.~\ref{fig2} (left) for the $m_h^{\rm max}$ scenario.  
The LHC is surprisingly sensitive: with high luminosity,
$h$ can be distinguished from the SM Higgs at the $5\sigma$ level in 
this scenario out to $M_A \simeq 350$ GeV.  The sensitivity
comes mostly from the WBF channels, as shown in Fig.~\ref{fig2} (right).
This shows the importance of trying hard to get the WBF channels to work
at high luminosity.

\section{Acknowledgements}

I thank M.~D\"uhrssen, S.~Heinemeyer, D.~Rainwater, G.~Weiglein
and D.~Zeppenfeld for a fruitful collaboration leading to the paper
\cite{us} on which this talk was based.
This work was supported in part by the
U.S.~Department of Energy under grant DE-FG02-95ER40896 and in part by
the Wisconsin Alumni Research Foundation.

\bibliographystyle{plain}

\end{document}